\documentclass[11pt]{article}

\usepackage[final]{acl}

\usepackage{times}
\usepackage{latexsym}
\usepackage{setspace}
\usepackage[T1]{fontenc}

\usepackage{minted}

\usepackage{caption}
\usepackage[most]{tcolorbox}
\tcbuselibrary{breakable, skins}
\usepackage[utf8]{inputenc}
\usepackage[ruled,linesnumbered]{algorithm2e}
\usepackage{algpseudocode}
\usepackage{microtype}

\usepackage{inconsolata}
\usepackage{enumitem} 
\usepackage{amsmath}
\usepackage{amssymb} 
\usepackage{booktabs}
\usepackage{multirow}
\usepackage{float}
\usepackage[table,xcdraw]{xcolor}
\usepackage{colortbl,pgf}
\usepackage{listings}
\usepackage{newfloat}
\usepackage{graphicx}
\usepackage{tikz}          %
\usepackage{url}

\newcommand{\framework}{\texttt{ARK}}
\usepackage{xfp} %

\newcommand{\scorecell}[1]{%
  \begingroup
  \edef\colval{\fpeval{max(0, min(100, round(#1-50)))}}%
  \ifnum\colval>0\relax
    \edef\doCell{\noexpand\cellcolor{blue!\colval}}%
    \doCell #1%
  \else
    \cellcolor{white}#1%
  \fi
  \endgroup
}

\title{ARK: Answer-Centric Retriever Tuning via \\ KG-augmented Curriculum Learning}

\author{
  Hang Ding \textsuperscript{1}\thanks{Equal contribution. Author order is randomized.},
  Jiawei Zhou \textsuperscript{1}\footnotemark[1],
  Haiyun Jiang \textsuperscript{1}\thanks{Corresponding Author.} \\
  \textsuperscript{1} Shanghai Jiao Tong University \\
  \texttt{\{dearsloth,davidzjw,haiyun2025\}@sjtu.edu.cn}
}

\begin{document}
\maketitle
\begin{abstract}
Retrieval-Augmented Generation (RAG) has emerged as a powerful framework for knowledge-intensive tasks, yet its effectiveness in long-context scenarios is often bottlenecked by the retriever's inability to distinguish sparse yet crucial evidence. Standard retrievers, optimized for query-document similarity, frequently fail to align with the downstream goal of generating a precise answer. To bridge this gap, we propose a novel fine-tuning framework that optimizes the retriever for Answer Alignment. Specifically, we first identify high-quality positive chunks by evaluating their sufficiency to generate the correct answer. We then employ a curriculum-based contrastive learning scheme to fine-tune the retriever. This curriculum leverages LLM-constructed Knowledge Graphs (KGs) to generate augmented queries, which in turn mine progressively challenging hard negatives. This process trains the retriever to distinguish the answer-sufficient positive chunks from these nuanced distractors, enhancing its generalization. Extensive experiments on 10 datasets from the Ultradomain and LongBench benchmarks demonstrate that our fine-tuned retriever achieves state-of-the-art performance, improving 14.5\% over the base model without substantial architectural modifications and maintaining strong efficiency for long-context RAG. Our work presents a robust and effective methodology for building truly answer-centric retrievers. Source Code is available on \url{https://github.com/valleysprings/ARK/}.
\end{abstract}

\section{Introduction}

Large Language Models (LLMs) have achieved human-level performance on many NLP tasks \citep{achiam2023gpt,touvron2023llama}, but still struggle with long-term memory, often omitting or conflating details in scenarios requiring complex reasoning or extended context \citep{li2024long,lazaridou2021mind,su2025racqc}. 

Retrieval-Augmented Generation (RAG) \citep{lewis2020retrieval} addresses this limitation by connecting LLMs to external knowledge sources, refreshing their memory at inference time. Since its introduction, RAG has rapidly evolved from naive RAG pipeline, with vector retrieval \citep{shi2023replug,borgeaud2022improving} to advanced pipelines that incorporate recursive chunking \citep{sarthi_raptor_2024}, knowledge graphs (KGs) \citep{edge2024local,chang2025grail}, and internal memory modules \citep{qian2024memorag}, substantially improving the handling of long-context input.

While KG-integrated pipelines \citep{edge2024local} have shown promising gains for complex summarization, they suffer from efficiency and accuracy bottlenecks in broader retrieval tasks. The indexing phase in KG-based RAG \citep{edge2024local} and follow-up works \citep{gutierrez2024hipporag,guo2025lightragsimplefastretrievalaugmented} requires processing extremely large token volumes with powerful LLMs, resulting in high computational cost. In addition, KGs often struggle with fine-grained entity disambiguation: community-curated clusters, though rich, are noisy and insufficiently filtered. Consequently, retrieval may aggregate irrelevant or even conflicting content, reducing both the consistency and quality of generated outputs.

To train a retriever for true answer sufficiency, we propose \framework{} (\underline{\textbf{A}}nswer-centric \underline{\textbf{R}}etriever fine-tuning via \underline{\textbf{K}}G-driven curriculum), a framework that redefines the role of Knowledge Graphs in RAG. Rather than serving as a direct retrieval source, the KG powers an \textbf{Answer-Centric Curriculum Learning} scheme, enabling fine-grained discrimination, sufficiency-aware retrieval, and improved generalization of the retriever.

At its core, \framework{} first identifies what makes evidence truly useful—whether it suffices to generate the correct answer. We formalize this with an in-context sufficiency metric combining three alignment strategies (Forward, Backward, Retriever) to extract high-quality \textbf{positive chunks} as anchors. Building on these positives, \framework{} leverages the KG as a \textbf{hard-negative generator}: It runs Personalized PageRank(PPR) over the KG to extract answer-relevant subgraphs, which in turn guide the creation of augmented queries. These queries are specifically designed to mine progressively challenging hard negatives. Concretely, we use PPR on a query-specific subgraph to expose co-occurrence neighbors near the gold entities, and inject them into query augmentations so the retriever is drawn to false positives—highly related yet insufficient evidence. This community-aware mining yields harder, more calibration-relevant negatives than random or keyword-based baselines. Through contrastive training against this curriculum, \framework{} learns to prioritize truly answer-informative segments while filtering misleading context, mastering both sufficiency and fine-grained discrimination.

\textbf{To summarize, our contributions are}:
\begin{itemize}[leftmargin=*, itemsep=0pt, topsep=0pt, parsep=0pt]
    \item We propose \framework, a framework that finetunes the retriever through contrastive learning for scalable long-context retrieval.
    \item We devise a synthetic query-generation pipeline that uses KG subgraphs to produce challenging hard negatives, and integrate them into an answer-centric curriculum learning scheme that progressively increases negative difficulty.
    \item We introduce an in-context answer sufficiency metric, composed of three complementary alignment scores (Forward, Backward, and Retriever), to identify high-quality positive chunks that serve as the anchor for contrastive learning.
    \item Through extensive experiments, we demonstrate that \framework{} achieves state-of-the-art retrieval performance on 8 out of 10 datasets across the LongBench and Ultradomain benchmarks, with an average F1-score improvement of 14.5\% over the base model, showcasing its effectiveness and efficiency.
\end{itemize}

\label{sec:introduction}

\section{Related work}

\subsection{Traditional RAG Techniques}

RAG systems enhance LLM outputs by combining retrieval with generation. Early implementations used fixed retrievers to supply documents to a reader model: classical methods such as BM25 \citep{robertson2009probabilistic} relied on lexical matching, while neural retrievers like DPR \citep{karpukhin2020dense} employed dense embeddings for semantic retrieval. Typically, retrievers were trained on QA pairs, with readers fine-tuned independently.

Later work emphasized tighter integration. RAG models \citep{lewis2020retrieval} enabled end-to-end fine-tuning by treating retrieved documents as latent variables, aligning the retriever with a BART-based generator. This improved recall and consistency but increased complexity due to non-differentiable retrieval. To mitigate this, techniques such as hard negative mining \citep{robinson2020contrastive} and knowledge distillation from generators into retrievers \citep{izacard2021distilling} were proposed, further enhancing retriever–reader synergy.

\subsection{Advanced RAG Techniques}

Recent advances in RAG aim to improve retrieval reasoning and integrate hybrid knowledge sources. One line of work focuses on query rewriting and decomposition. RQ-RAG \citep{chan2024rq} enhances multi-hop QA by decomposing complex queries into simpler sub-queries, while HyDE \citep{gao2022precisezeroshotdenseretrieval} generates hypothetical documents that serve as refined queries to boost retrieval precision. Context rewriting techniques have also emerged. MemoRAG \citep{qian2024memorag} further addresses long-context retrieval by compressing memory and producing clue phrases to guide retrieval, composing answers from retrieved snippets.
\begin{figure*}[!t]
  \centering
  \includegraphics[width=\textwidth]{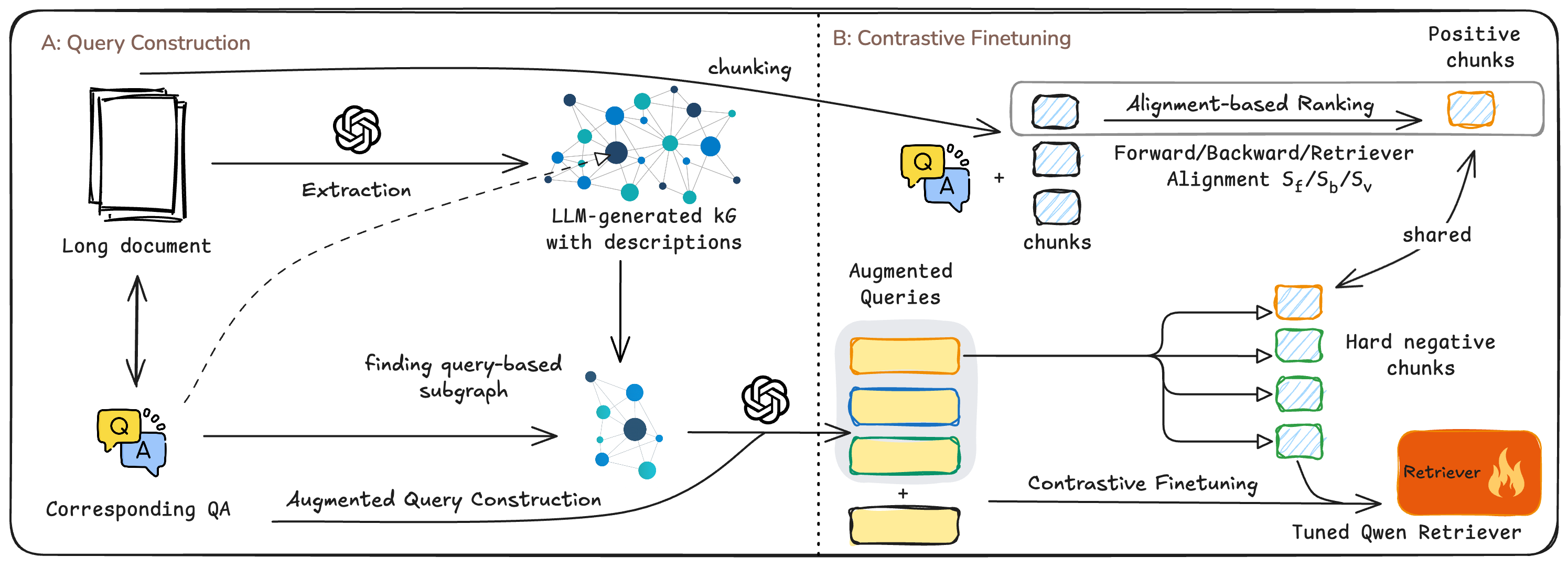}
  \caption{\textbf{Our RAG Retriever Finetuning Framework }\framework, which consists of two major stages: A (Query Construction): From long documents and their corresponding QA pairs, we extract a query-based subgraph using an LLM-generated KG. The subgraph is reformulated with knowledge injection to produce enriched queries. B (Contrastive Finetuning): Using both the original query and injected variants, we identify positive chunks (via alignment scoring) and hard negatives (that match injected queries but differ semantically from ground truth). }
  \label{fig:framework}
  \vskip -0.1in
\end{figure*}

Another direction integrates symbolic and neural approaches, often by incorporating KGs into retrieval pipelines. GraphRAG \citep{edge2024local} constructs graphs from documents and summarizes clusters into “community reports” via community detection, supporting improved multi-hop reasoning. LightRAG \citep{guo2025lightragsimplefastretrievalaugmented} simplifies this pipeline by first retrieving low-level nodes and then following graph links to higher-level concepts, improving both recall and efficiency. HippoRAG \citep{gutierrez2024hipporag} models memory consolidation using KGs and PPR, retrieving subgraphs at query time to emulate long-term memory access. Furthermore, many other RAG methodologies\citep{ding-etal-2026-scirag,chen2026grorag,liu2026discoragdiscourseawareretrievalaugmentedgeneration,xu2026unlocking,li2026sesearchselfevolvingsearchagent} also collectively contribute to the development of RAG capabilities.

\section{Methodology}

\subsection{Framework}
Our proposed framework, \framework{}, follows a two-stage architecture—\textbf{Query Construction} and \textbf{Contrastive Finetuning}—as illustrated in Figure~\ref{fig:framework}. The first stage prepares curriculum components, while the second performs training.

In \textbf{Query Construction}, we build an LLM-derived KG from the context, identify entities from the QA pair, and extract a query-specific subgraph. This subgraph supports the creation of \textit{injected queries}, which preserve the semantics of the original while adding contextual structure. These enriched queries serve as the basis for generating hard negatives.
\begin{figure}[ht!]
  \centering
  \includegraphics[width=\columnwidth]{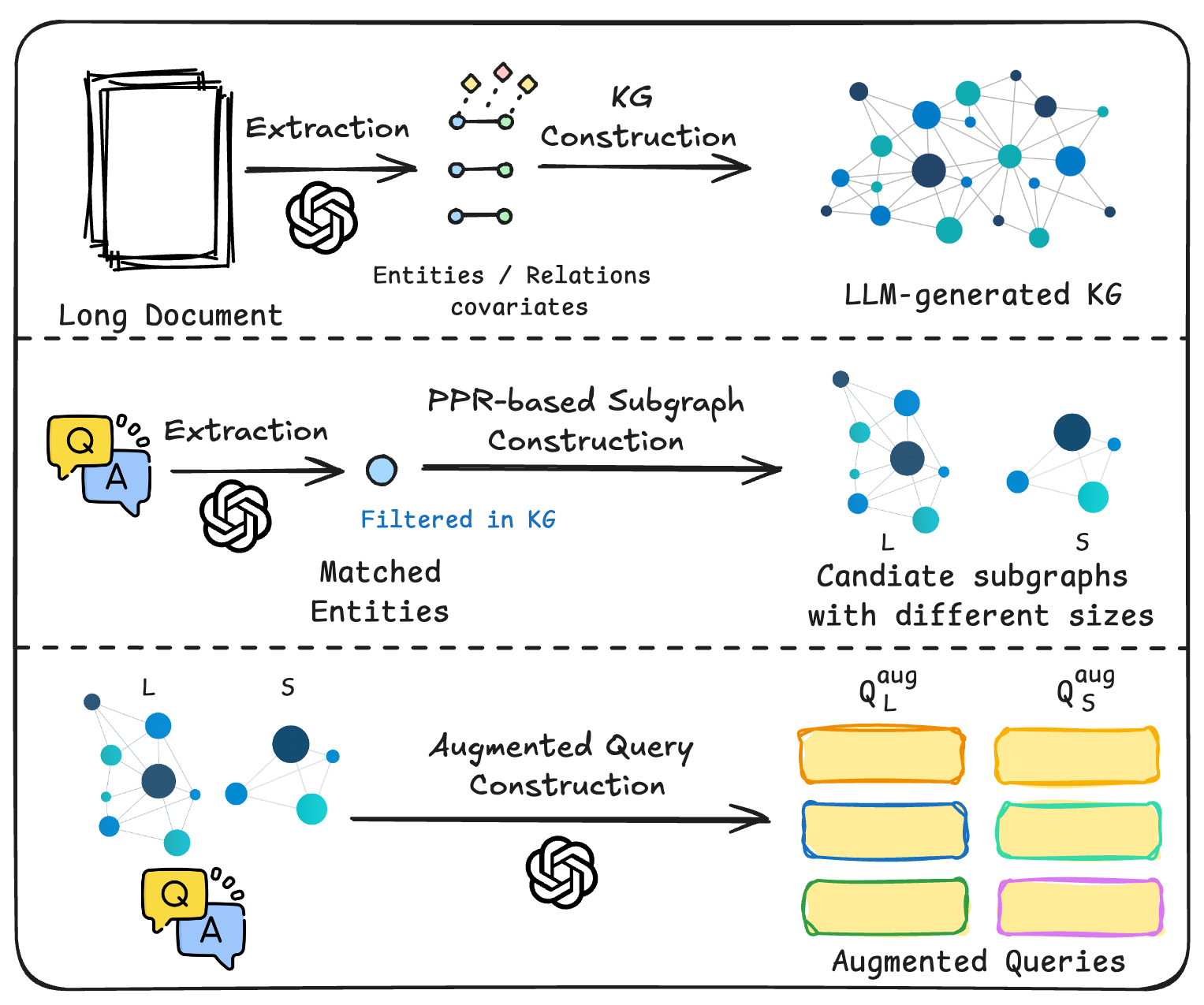}
  \caption{\textbf{Query Construction Phase.} The pipeline begins with \emph{KG Construction}, where we extract entities, relations, and covariates from long documents to construct an LLM-generated KG. Given a corresponding QA pair, relevant entities are extracted and used to \emph{construct PPR-based subgraphs} from the KG, with varying maximum sizes to control difficulty. Finally, \emph{Augmented Queries} are formulated with LLM conditioned on these candidate subgraphs.}
  \label{fig:data_gen}
  \vskip -0.1in
\end{figure}

In \textbf{Contrastive Finetuning}, we define positives using our in-context answer sufficiency metric (combining three alignment scores) to rank context chunks and select the most sufficient ones. Hard negatives are chunks that score highly for an injected query but are absent from the positive set. With these positives and negatives, we finetune the retriever via contrastive learning, enhancing its discriminative ability. The resulting retriever integrates seamlessly into existing RAG pipelines without architectural changes.

\subsection{KG-based Query Construction}\label{subsec:methodology1}

During training, to effectively extract high-quality, answer-guided queries from ultra-long source contexts, we develop an innovative pipeline that integrates LLM-assisted KG construction, PPR-based subgraph construction, and finally query formation. As illustrated in Figure \ref{fig:data_gen}, the data generation process consists of three steps.

\subsubsection{KG construction}

We adopt a lightweight prompt to extract entities and relations from fixed-size text chunks. Similar to GraphRAG \cite{edge2024local}, extraction is performed independently on each chunk. However, unlike GraphRAG, which constructs hierarchical summaries over the entire graph, we focus solely on entity–relation extraction without building a hierarchical index. This design is enabled by our answer-sufficient community construction mechanism, which removes the need for global graph-level indexing. As a result, we significantly reduce unnecessary LLM calls and improve computational efficiency.

In light of the inherent locality bias of LLMs, which primarily operate within limited context windows, the resulting KG may exhibit lower edge density compared to traditional KGs, since only intra-chunk relationships are explicitly extracted. To mitigate this sparsity, we further augment the graph by introducing undirected edges between entity pairs whose embedding similarity exceeds a predefined threshold \(\tau\). This similarity-based augmentation enhances graph connectivity and supports more effective downstream retrieval.

\subsubsection{Subgraph Extraction} To capture neighborhoods most relevant to the input answer query, we employ community search, which identifies query-dependent subgraphs rather than global clusters. This aligns with our retrieval objective: surfacing context that is semantically close to the query yet hard to distinguish from true positives, thus serving as high-quality hard negatives. LLM-derived knowledge graphs are typically sparse, with edges concentrated within intra-chunk links, making purely topology-driven methods less effective. To address this, we adopt Personalized PageRank (PPR) seeded on query entities to extract semantically coherent communities. 

\begin{algorithm}[!t]
\small
\setstretch{1.1}
\caption{Cohesive Subgraph Extraction}
\label{alg:train-data-generation}
\begin{algorithmic}[1]
\Require Matched entities $V_a$, LLM-generated KG $G(V,E)$, PPR parameter $\alpha,\epsilon$, subgraph size threshold $k_{\min},k_{\max}$
\Ensure Answer-related entities $V_\texttt{comm}$
\State $\chi_{a} \gets$ Indicator function of $V_a$
\State Compute $\texttt{apr}_{\texttt{comm}}(\alpha, \chi_{a})$ with $\epsilon$ threshold
\State $\{\texttt{apr}_i, v_i\}_{i\in V} \gets \texttt{sort}(\texttt{apr}_{\texttt{comm}}, \texttt{desc} = \texttt{True})$
\State $\{\texttt{apr}_i, v_i\}_{i\in V^\prime} \gets$ Filter trailing terms of $\{\texttt{apr}_i,v_i\}_{i\in V}$ where $\{i \mid \texttt{apr}_i < \epsilon\}$
\State $\{\texttt{apr}_i, v_i\}_{i\in V^\prime} \gets \{-\log \texttt{apr}_i, v_i\}_{i\in V^\prime}$
\State $\texttt{index} \gets \Delta_{\arg\max_{\{i| i \leq \min \{k_{\max}, |V^\prime|\},i\geq k_{\min}\}}} \texttt{apr}$
\State $V_\texttt{comm} \gets$ First $\texttt{index}$ entities from $\{\texttt{apr}_i, v_i\}_{i\in V^\prime}$
\end{algorithmic}
\end{algorithm}

However, due to the sparsity of LLM-derived KGs, where edges are primarily confined to intra-chunk relationships, conventional topology-driven community search methods often fail to perform effectively. To overcome this limitation, we utilize PPR to assess and construct communities based on semantically aligned entities.
We begin by extracting the positive query entities \(V_a\) from the answer and identifying their corresponding entities within the KG. Using these matched entities, we perform PPR based on the matched ones, formally defined as:
\begin{equation}
\operatorname{pr}(\alpha, \chi_{a})=\alpha \chi_a +(1-\alpha) W\operatorname{pr}(\alpha, \chi_a) 
\end{equation}
\noindent Here, \(\operatorname{pr}(\alpha, \chi_{a})\) denotes the PPR value for the given entities \(a\) on KG \(G\), \(\chi_a\) denotes the indicator function of \(a\), \(\alpha\) denotes the teleport probability, and \(W = A^\top D^{-1}\) represents the normalized transition matrix based on adjacency matrix \(A\) and degree matrix \(D\) from graph \(G\). 

Due to scalability challenges in computing PPR over large graphs, we employ power iteration to approximate the solution. The approximation vector \(\operatorname{apr}^{(0)}(\alpha, \chi_a)\) is initialized as \(\chi_a\), and iteratively updated as follows:
\begin{equation}
\operatorname{apr}^{(n+1)}(\alpha, \chi_a)=\alpha \chi_a+(1-\alpha) W\operatorname{apr}^{(n)}(\alpha, \chi_a) 
\end{equation}

Given the absolute approximation error bound: \(\left\|\operatorname{apr}^{(n+1)}(\alpha, \chi_a)-\operatorname{apr}^{(n)}(\alpha, \chi_a)\right\|_{\infty} < \epsilon\) The computational complexity of the iterative method is \(O\left(|E| \log \frac{1}{\epsilon}\right)\), as initially proposed by \cite{haveliwala2003second}. 

Following the approach of \citep{andersen2007detecting, zhou2025comet}, we detect sharp drops in the approximated PPR scores to delineate topologically coherent communities, but here we choose the largest first difference of $|\log \texttt{apr}|$ as it naturally shapes a cohesive subgraph. From the resulting subgraph, community information is subsequently used to construct augmented queries to generate hard negatives $\mathcal{T}^-_{\texttt{hard}}$ used for the retriever. The overall subgraph extraction is outlined in Algorithm~\ref{alg:train-data-generation}.

\begin{figure*}[t]
    \centering
    \includegraphics[width=\textwidth]{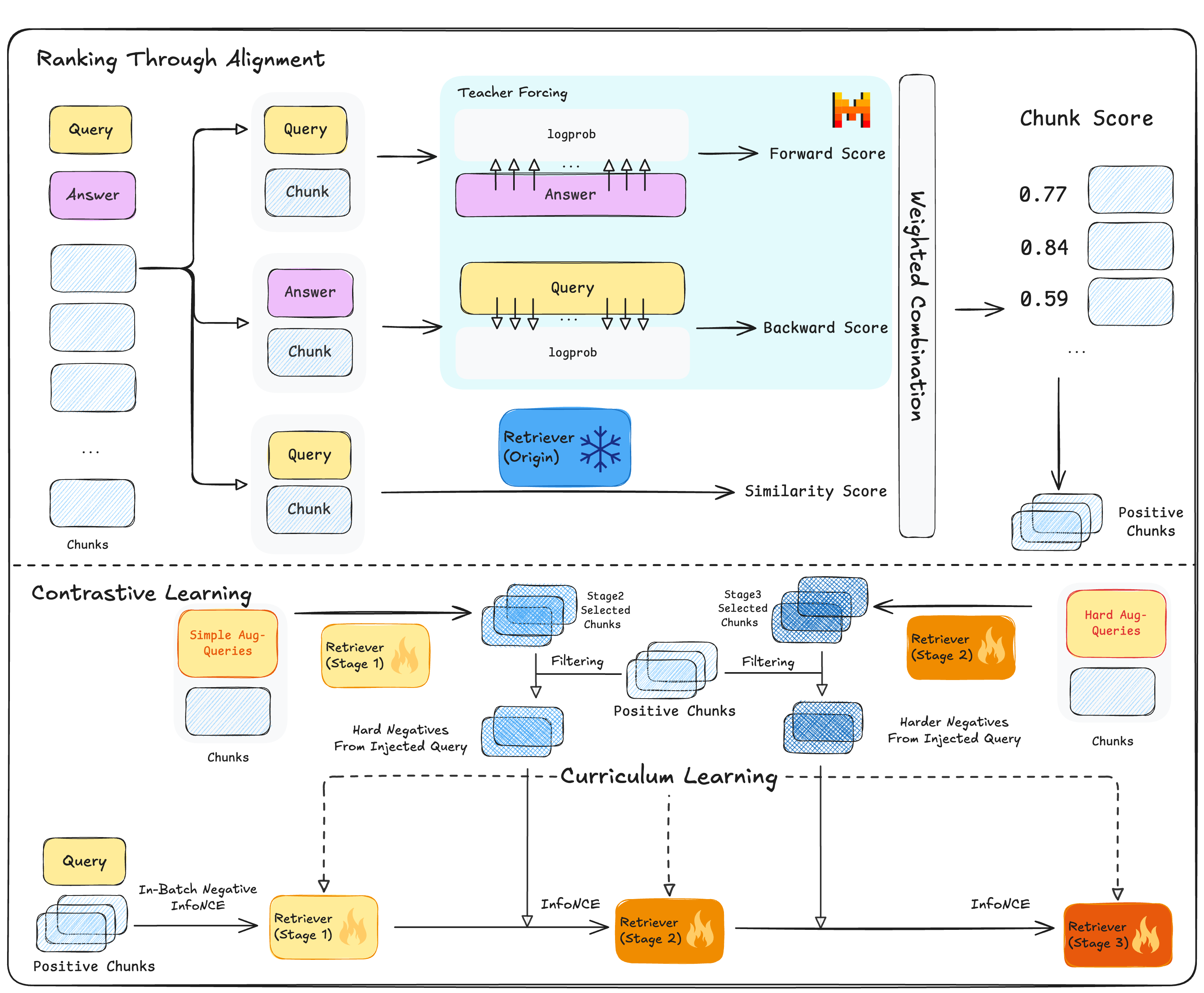}
    \caption{\textbf{Contrastive Finetuning Phase.} Our fine-tuning pipeline comprises two sequential components: \emph{Ranking Alignment}, in which for each sample, we combine three alignment scores to select the Top-\(M\) chunks as positive chunks; followed by \emph{Curriculum-based Contrastive Learning}, which progressively refines the retriever through (i) in-batch negative sampling, (ii) hard negatives $\mathcal{T}^-_{\texttt{hard}_L}$ mined via query set $\mathcal{Q}^{\texttt{aug}}_{L}$, and (iii) more challenging negatives $\mathcal{T}^-_{\texttt{hard}_S}$ obtained from $\mathcal{Q}^{\texttt{aug}}_{S}$.}
    \label{fig:FT}
    \vskip -0.15in
\end{figure*}

\subsubsection{Query Formation}

Once we have retrieved community entities $V_\texttt{comm}$, we synthesize a richer query collection $\mathcal{Q}^{\texttt{aug}}$ by perturbing and extending the original $\texttt{QA}$ instance. Each new query carries additional context, such as answer spans, related entities, or semantic relations, and thus provides the retriever with more signals for hard negative chunks. Concretely, for each original QA pair, we sample subgraph entities $V_\texttt{comm} \subseteq V$ of controllable size and connectivity, and then apply the query‐transformation prompts to generated the given set of augmented query $\mathcal{Q}^{\texttt{aug}}$.

We instantiate each augmented query $\mathcal{Q}^{\texttt{aug}}$ based on external LLM over subgraphs of two different sizes. Here, we construct a diverse pool of augmented queries $\mathcal{Q}^{\texttt{aug}} = \{q_{1}, \dots, q_{N}\}$, categorized into large and small variants, $\mathcal{Q}^{\texttt{aug}}_{L}$ and $\mathcal{Q}^{\texttt{aug}}_{S}$, respectively. Each query type is paired with a corresponding set of hard negatives, denoted as $\mathcal{T}^-_{\texttt{hard}_L}$ and $\mathcal{T}^-_{\texttt{hard}_S}$. This augmentation strategy facilitates the curriculum learning of the retriever, which enhances its capacity to capture fine-grained semantic distinctions.

\subsection{Alignment-Based Fine-Tuning}\label{subsec:methodology2}

After extracting answer-relevant chunks, we fine-tune the retriever using a multi-stage framework. Specifically, we introduce an answer-centric in-context scoring approach, carefully designed to mitigate weight collapse by maintaining balanced and informative gradients. Building upon this alignment-based scoring and ranking, we progressively apply a curriculum learning strategy that incrementally exposes the retriever to increasingly challenging negative samples, thereby systematically enhancing its discriminative power and ensuring stable generalization performance.

\subsubsection{Alignment-Based Ranking}
To retrieve positive chunks, we propose several alignment-based scoring functions that link a candidate chunk not only to the query but also to the expected answer, which is relevant to the query and sufficient to support accurate answer generation.

\paragraph{Forward Alignment}
Given a chunk \(t\) and the full question \(q\), we evaluate the likelihood of the generator LLM (parameterized by \(\theta\)) reproducing the reference answer \(a=\langle a_{1},\dots ,a_{|a|}\rangle\). This score quantifies the \textit{sufficiency} of the chunk \(t\) to generate the correct answer. During scoring, the concatenated query and chunk prompt \([q;t]\) is fed to the model. We employ teacher forcing, using the ground-truth answer tokens as targets to calculate the mean token log-likelihood without them being part of the conditioning context. The forward alignment score \(S_f\) is thus defined as:
\begin{equation}
S_{f}(q,t,a)=\frac{1}{|a|}\sum_{i=1}^{|a|}\log p_{\theta}(a_{i}|q,t,a_{<i})
\label{eq:forward_alignment}
\end{equation}
This score is computed in a single forward pass by aggregating the log-softmax probabilities for each token in the ground-truth answer. 

\paragraph{Backward Alignment}
Analogously, to measure the \textit{relevance} of a chunk in linking the answer back to the original query, we pair the chunk \(t\) with the answer \(a\) and task the model with reconstructing the question \(q\). The backward score \(S_b\) is calculated using the same teacher-forcing technique:
\begin{equation}
S_{b}(a,t,q)=\frac{1}{|q|}\sum_{j=1}^{|q|}\log p_{\theta}(q_{j}|a,t,q_{<j})
\label{eq:backward_alignment}
\end{equation}
This bidirectional scoring mechanism ensures that selected chunks are strongly correlated with the reasoning path from question to answer. 

\paragraph{Parameter Alignment}
To regularize the fine-tuning process and mitigate catastrophic forgetting, we incorporate the original retriever's similarity score, \(S_v\), as a form of parameter alignment. This score uses the cosine similarity to preserve the geometric structure learned inherent in the original retriever.
\begin{equation}
S_{v}(q,t)=\operatorname{sim}(q,t)
\label{eq:parameter_alignment}
\end{equation}
The final unified score \(S\) for a given chunk \(t\) and QA \(q,a\) is a weighted combination of these three components. 
We did not tune these weights, as we believe adjustment is unnecessary. 
Instead, we adopt intuitive fixed values: equal weights for forward and backward alignment, and a slightly lower weight for parameter regularization (\(\lambda_f=1.0\), \(\lambda_b=0.3\), \(\lambda_v=1.0\)).
\begin{equation}
S(t) = \lambda_{f}S_{f} + \lambda_{b}S_{b} + \lambda_{v}S_{v}
\label{eq:combined_score}
\end{equation}
This unified score serves as the primary criterion for identifying high-quality positive samples for the initial stage of our fine-tuning curriculum.

\subsubsection{Curriculum-Based Contrastive Finetuning}
Illustrated in Figure \ref{fig:FT}. We structure the fine-tuning process as a three-stage curriculum, where the difficulty of the discrimination task increases at each stage. This approach allows the model to first learn a robust answer-centric representation and then progressively refine it by focusing on increasingly subtle and challenging distractors.

\paragraph{Stage 1: Initial Answer-Centric Alignment}
The primary goal of this stage is to align the retriever with chunks that are highly conducive to generating the correct answer. For each query \(q\), we calculate the unified score \(S(t)\) for all candidate chunks in \(\mathcal{T}\). We select the top-$M$ chunks with the highest scores as the positive set \(\mathcal{T}^+\). The retriever is then trained using a contrastive objective. Specifically, for a given positive pair \((q, t^+)\) where \(t^+ \in \mathcal{T}^+\), we use other positive chunks within the same mini-batch as in-batch negatives. This is an effective and efficient method for initial training. The loss for this stage is the InfoNCE loss:
\begin{equation}
\mathcal{L}_{\text{Stage1}} = -\log \frac{\exp(\operatorname{sim}(q, t^+)/\tau)}{\sum_{t'_j \in \text{Batch}} \exp(\operatorname{sim}(q, t'_j)/\tau)}
\end{equation}
where \(\operatorname{sim}(\cdot,\cdot)\) is the cosine similarity from the retriever being trained and \(\tau\) is a temperature hyperparameter.

\paragraph{Stage 2: Coarse Alignment with $\mathcal{Q}^{\texttt{aug}}_L$}
After the initial alignment, we further enhance the retriever's robustness by incorporating hard negatives. Specifically, we leverage $\mathcal{Q}^{\texttt{aug}}_L$ generated from the KG. Using the retriever fine-tuned in Stage 1, we retrieve the top-$K$ chunks for each complex query \(q\in \mathcal{Q}^{\texttt{aug}}_L\). From this set, we exclude any chunks that appear in the ground-truth positive set \(\mathcal{T}^+\). The remaining chunks constitute the hard negative set \(\mathcal{T}^-_{\texttt{hard}_L}\). Compared to \(\mathcal{T}^-_{\texttt{hard}_S}\), these negatives are less challenging due to their greater semantic diversity. The retriever is then further trained to distinguish the original positive chunks from these hard negatives.
\begin{equation}
\mathcal{L}_{\text{Stage2}}
= -\log \frac{e^{s(q,t^+)}}{\sum_{t\in \mathcal{C}(q)} e^{s(q,t)}} ,
\end{equation}
\noindent\textit{where} $s(q,t)=\operatorname{sim}(q,t)/\tau$ and $\mathcal{C}(q)=\{t^+\}\cup \mathcal{T}^-_{\text{hard}_L}$.

\paragraph{Stage 3: Fine-Grained Alignment with $\mathcal{Q}^{\texttt{aug}}_S$}
In the final stage, we further sharpen the retriever using the simple augmented queries $\mathcal{Q}^{\texttt{aug}}_S$, which are minor perturbations of the original query (e.g., with only distracting covariates added). We use the retriever from Stage 2 to retrieve chunks and, after filtering out the golden positives, obtain a set of "harder" negative chunks \(\mathcal{T}^-_{\texttt{hard}_S}\). The training objective remains the InfoNCE loss, but with a more challenging target.

\section{Experiment}\label{sec:experiment}

\begin{table*}[ht]
  \centering
  \small
  \caption{\textbf{Main Evaulation results}. The evaluation metrics are F1-score / Win Rate (\%), with the \textcolor{red}{\textbf{best}} results highlighted in bold and the \textcolor{blue}{\underline{second-best}} results underlined. The improvement rate (↑ \%, (ARK- Base) / Base) is calculated based on our base model Qwen3-embedding. Cell shading indicates relative win rates compared to \framework.}
  \label{tab:performance1}
  \resizebox{\textwidth}{!}{%
  \begin{tabular}{cccccccccccc}
    \toprule
    \multirow{2}{*}{\textsc{\textbf{Metrics}}} & \multirow{2}{*}{\textsc{\textbf{Models}}}
      & \multicolumn{5}{c}{\textbf{LongBench}}
      & \multicolumn{5}{c}{\textbf{UltraDomain}} \\
    \cmidrule(lr){3-7}  \cmidrule(lr){8-12}
    & & \textbf{nar} & \textbf{qas} & \textbf{mus} & \textbf{2wiki} & \textbf{hot}
      & \textbf{bio} & \textbf{fic} & \textbf{music} & \textbf{tech} & \textbf{phil} \\
    
    \midrule
    \multirow{11}{*}{\textsc{\textbf{F1}}} &
    Full
      & 12.95 & 22.79 & 6.74  & 20.13 & 26.87 & 27.47 & 25.75 & 25.50 & 22.68 & 23.05 \\
    \cmidrule(lr){2-12}
    & Qwen3-embedding
      & 19.58 & \underline{\textcolor{blue}{23.90}} & 14.19 & 21.24 & 35.27 & 32.99 & 29.41 & 34.90 & 38.03 & 34.04\\
    & BGE-M3
      & 18.37 & 23.33 & \textbf{\textcolor{red}{21.13}} & 22.86 & 38.64 & 32.52 & 31.72 & \underline{\textcolor{blue}{35.34}} & 39.13 & 35.97 \\
    & Stella-v5
      & \underline{\textcolor{blue}{20.90}} & 23.39 & 17.08 & 22.13 & 35.45 & 33.85 & \underline{\textcolor{blue}{32.41}} & 35.02 & 35.16 & 34.09 \\
    & Jina-emb-v3
      & 19.39 & 20.70 & 20.58 & 19.34 & \underline{\textcolor{blue}{39.17}} & 32.88 & 29.00 & 33.74 & 38.74 & \underline{\textcolor{blue}{36.81}} \\
    \cmidrule(lr){2-12}
    & GraphRAG
      & 4.21  & 7.69  & 2.15  & 5.52  & 3.03  & 18.87 & 16.92 & 14.97 & 21.93 & 20.01\\
    & LightRAG
      & 2.65 & 3.25 & 1.95 & 3.67 & 2.74 & 16.06 & 14.13 & 15.08 & 12.19 & 14.04\\
    & HippoRAG
      & 11.51 & 21.90 & 13.09 & \textbf{\textcolor{red}{30.96}} & 28.71 & \underline{\textcolor{blue}{36.13}} & 29.23 & 32.94 & 27.15 & 29.06 \\
    & MemoRAG
      & 15.49 & 17.96 & 8.74  & 16.57 & 22.79 & 31.08 & 27.87 & 33.26 & \underline{\textcolor{blue}{39.14}} & 31.98 \\
    \cmidrule(lr){2-12}
    & \textbf{\framework(Ours)} 
      & \textbf{\textcolor{red}{21.57}} & \textbf{\textcolor{red}{24.04}} & \underline{\textcolor{blue}{20.60}} & \underline{\textcolor{blue}{23.41}} & \textbf{\textcolor{red}{42.35}} & \textbf{\textcolor{red}{36.19}} & \textbf{\textcolor{red}{32.59}} & \textbf{\textcolor{red}{38.03}} & \textbf{\textcolor{red}{40.16}} & \textbf{\textcolor{red}{37.86}} \\
    
    & $\uparrow\% $ & ${10.16}$& ${0.59}$& ${45.17}$& ${10.22}$ & ${20.07}$& ${9.70}$& ${10.81}$& ${8.97}$& ${5.60}$& ${11.22}$\\

    \midrule

        \multirow{9}{*}{\textsc{\textbf{Win rate of ARK}}} &
    Full
      & \scorecell{83.33}& \scorecell{46.03}& \scorecell{80.00}& \scorecell{64.52}& \scorecell{70.97}& \scorecell{95.00} & \scorecell{100.00} & \scorecell{88.89} & \scorecell{94.74} & \scorecell{100.00} \\
    \cmidrule(lr){2-12}
    & Qwen3-embedding
      & \scorecell{58.33} & \scorecell{52.54} & \scorecell{63.46} & \scorecell{57.14} & \scorecell{68.89}
      & \scorecell{95.00} & \scorecell{85.71} & \scorecell{94.74} & \scorecell{78.57} & \scorecell{55.56} \\
    & BGE-M3
      & \scorecell{60.00} & \scorecell{50.77} & \scorecell{56.00} & \scorecell{52.54} & \scorecell{52.83}
      & \scorecell{70.59} & \scorecell{84.62} & \scorecell{58.82} & \scorecell{73.33} & \scorecell{60.00} \\
    & Stella-v5
      & \scorecell{65.67} & \scorecell{66.67} & \scorecell{58.00} & \scorecell{50.00} & \scorecell{67.39}
      & \scorecell{72.22} & \scorecell{62.50} & \scorecell{64.71} & \scorecell{71.43} & \scorecell{89.47} \\
    & Jina-emb-v3
      & \scorecell{63.08} & \scorecell{54.84} & \scorecell{54.90} & \scorecell{57.41} & \scorecell{43.24}
      & \scorecell{77.78} & \scorecell{61.54} & \scorecell{66.67} & \scorecell{58.82} & \scorecell{50.00} \\
    \cmidrule(lr){2-12}
    & GraphRAG
      & \scorecell{93.62} & \scorecell{90.70} & \scorecell{78.26} & \scorecell{83.75} & \scorecell{78.57}
      & \scorecell{100.00}& \scorecell{100.00}& \scorecell{85.00} & \scorecell{95.00} & \scorecell{100.00}\\
    & LightRAG
      & \scorecell{96.63} & \scorecell{96.70} & \scorecell{91.46} & \scorecell{91.36} & \scorecell{96.74}
      & \scorecell{100.00}& \scorecell{100.00}& \scorecell{95.00} & \scorecell{100.00} & \scorecell{100.00}\\
    & HippoRAG
      & \scorecell{87.34}& \scorecell{53.85} & \scorecell{58.21} & \scorecell{34.67} & \scorecell{60.76}
      & \scorecell{77.78} & \scorecell{44.44} & \scorecell{62.50} & \scorecell{70.00} & \scorecell{70.00} \\
    & MemoRAG
      & \scorecell{80.00} & \scorecell{75.00} & \scorecell{66.22} & \scorecell{57.14} & \scorecell{67.65}
      & \scorecell{92.86} & \scorecell{94.12} & \scorecell{84.21} & \scorecell{87.50} & \scorecell{88.24} \\

    \bottomrule
  \end{tabular}%
  }
  \vskip -0.1in
\end{table*}

\subsection{Datasets} \label{sec:datasets}
In this paper, we focus on the long-text QA task. For training, we sample 200 cases each from the Finance and Legal domains (both from the Ultradomain dataset) to generate augmented queries. For evaluation, we first select five domains from the Ultradomain benchmark in MemoRAG \citep{qian2024memorag}, namely Biology, Fiction, Music, Technology, and Philosophy, each represented by a distinct domain-specific dataset.  We also use five LongBench \citep{bai2023longbench} dataset which includes both single-document QA: NarrativeQA \citep{kovcisky2018narrativeqa}, Qasper \citep{dasigi2021dataset} and multi-document QA: HotpotQA \citep{yang2018hotpotqa}, 2WikiMQA \citep{ho2020constructing},
and MuSiQue \citep{trivedi2022musique}.

\subsection{Experiment Setup}\label{sec:exp_setup}

\paragraph{Our Framework}

We adopt \textbf{Qwen3-embedding} (0.6B variant) \cite{qwen3embedding} as our base model for fine-tuning, due to its SOTA and scalable performance across a wide range of downstream tasks, and a tailored, tunable framework \texttt{SWIFT} \cite{zhao2024swiftascalablelightweightinfrastructure}. For alignment phase, we retrieve $10$ positive queries based on $S(t)$. For stage 2-3, we first sample respective $10$ augmented queries $\mathcal{Q}^{\texttt{aug}}_L/ \mathcal{Q}^{\texttt{aug}}_S$ per query based on the answer-aligned subgraph. Based on augmented queries, we sample the top $20$ retrieved chunks from $\mathcal{Q}^{\texttt{aug}}_L/ \mathcal{Q}^{\texttt{aug}}_S$ excluding any positive overlaps.

\paragraph{Model Selection} We first employ the latest \textbf{Gemini-2.5-Flash} \citep{comanici2025gemini} as our foundational model for OpenIE. The construction of our KG follows the GraphRAG \citep{edge2024local}, but the community generation is different. The NetworkX library is utilized to execute approximated PPR when conducting a query-based community search. We then employ \textbf{Gemini-2.5-Pro} \citep{comanici2025gemini} for constructing injected queries with the same ground truth as the original query. We use \textbf{GPT-4.1} to evaluate win rates via pairwise comparisons.

\paragraph{Baselines} We compare our approach against three categories of baselines: 1) \textbf{Full}: Directly providing the entire context to an LLM. 2) Dense Retrieval model: \textbf{Qwen3-embedding}: The original retriever without fine-tuning. \textbf{BGE-M3} \citep{bge_m3}: A hybrid retrieval model that integrates multiple strategies to achieve high accuracy and generalization across benchmarks. \textbf{Stella-v5} \citep{zhang2025jasperstelladistillationsota}: A top-ranking retriever on the MTEB leaderboard \citep{muennighoff2022mteb}. \textbf{Jina-emb-v3} \citep{sturua2024jinaembeddingsv3multilingualembeddingstask}: A powerful and widely-used multilingual, multi-task embedding model. 3) Advanced RAG methods: \textbf{GraphRAG} \citep{edge2024local}: Utilizes LLM-generated knowledge graphs and the Leiden algorithm for hierarchical retrieval. \textbf{LightRAG} \citep{guo2025lightragsimplefastretrievalaugmented}: Combines dual-level retrieval with vector and graph structures. \textbf{HippoRAG} \citep{gutierrez2024hipporag}: Leverages PPR for community-based retrieval.l. \textbf{MemoRAG} \citep{qian2024memorag}: Employs a lightweight long-context LLM to construct global memory and generate retrieval cues.

For online QA, we use \textbf{Mistral-7B-Instruct-v0.2}\citep{jiang2023mistral} as the default generator to mitigate potential pretraining contamination. Each context is split into 512-token chunks with 12-token overlaps, and the top-$5$ retrieved chunks are used for answer generation. We follow each dataset’s original evaluation protocol and report F1 score. For robustness, we also report the average pairwise win rate over five runs.

\subsubsection{Running Environment} \framework{} is implemented in Python 3.10 with PyTorch 2.7.1. All experiments are conducted on a machine equipped with 8 NVIDIA H20 GPU. We utilize \texttt{NetworkX}, \texttt{PyTorch}, \texttt{transformers}, and \texttt{sentence transformers} as the main libraries. In addition, we use \texttt{Ollama} for inference and \texttt{SWIFT} to finetune the Qwen3-embedding.

\subsection{Performance Evaluation}

Table~\ref{tab:performance1} reports the F1 scores and pairwise win-rate on the LongBench and UltraDomain benchmarks. Overall, \framework{} consistently outperforms its base model (Qwen3-embedding) across all datasets and achieves state-of-the-art performance compared to both KG-based baselines and top-ranking dense retrievers. In particular, \framework{} attains consistently higher pairwise win rates across the 10 evaluated datasets, outperforming both graph-based approaches (e.g., GraphRAG, LightRAG) and strong dense encoders (e.g., BGE-M3, Stella, Jina, Qwen3). Notably, it exceeds a 50\% win rate on the majority of benchmarks.

\framework{} shows strong generalization beyond its training domains (Finance and Legal), maintaining robust performance on unseen datasets, showing that our KG-guided curriculum not only improves retrieval accuracy but also enhances the ability to surface contextually meaningful evidence beyond the training distribution. Gains are especially pronounced on reasoning-intensive tasks such as MuSiQue and HotpotQA, where retrieval must synthesize dispersed evidence. By explicitly optimizing for \emph{answer sufficiency}, the retriever learns to favor chunks that are both relevant and sufficient for generating faithful answers. 
\begin{table}[htbp]
\small
\centering
\resizebox{1.05\columnwidth}{!}{
\begin{tabular}{llccccc}
\toprule
\textbf{\textsc{Stage}} & \textbf{\textsc{Scoring}} &  \textbf{mus}  &  \textbf{hot}  &  \textbf{bio}   &   \textbf{phil}  \\
\midrule
Original   &   -   &    14.19   &   35.27    &   32.99    &     34.04      \\
\midrule
\rowcolor{gray!10}
$1^\text{st}$  &   Full   &   \textcolor{blue}{18.44}    &    \textcolor{blue}{40.96}   &    \textcolor{blue}{32.35}   &        \textcolor{blue}{34.89}    \\
 & \quad w/o F.A. &   14.31    &   36.40    &   31.36    &       32.95     \\
 & \quad w/o B.A. &   16.73    &   39.74    &   33.57  &        34.12   \\
 & \quad w/o P.A. &  15.29    &   37.03    &   33.18    &       34.73     \\
 \rowcolor{gray!18}
 $2^\text{nd}$   &  Full &   \textcolor{blue}{18.47}    &   \textcolor{blue}{42.13}    &    \textcolor{blue}{34.74}  &        \textcolor{blue}{36.11}  \\     
\midrule
  \rowcolor{gray!30}
  $3^\text{rd}$ (\framework) & Full
  & \textcolor{blue}{\textbf{20.60}}   
  & \textcolor{blue}{\textbf{42.35}}   
  & \textcolor{blue}{\textbf{36.19}}    
  & \textcolor{blue}{\textbf{37.86}} \\
\bottomrule
\end{tabular}
}
\caption{\textbf{Ablation study}. \textsc{Stage} denotes finetuning stages (\textit{Original} is the base model), and \textsc{Scoring} specifies alignment types: Forward (F.A.), Backward (B.A.), and Parameter (P.A.).}
\label{tab:ablation}
\vskip -0.1in
\end{table}

On certain multi-hop tasks like 2Wiki, graph-centric methods such as HippoRAG remain competitive due to their traversal advantage. Nevertheless, \framework{} matches or surpasses their performance without requiring costly graph construction or long-context LLMs, thus offering higher efficiency. Moreover, unlike resource-intensive systems such as MemoRAG, which demand end-to-end training, our approach fine-tunes only the retriever. This modular design enables seamless integration into existing RAG pipelines, making \framework{} both practical and scalable. In summary, by emphasizing \emph{answer sufficiency} while preserving \emph{query similarity}, \framework{} consistently yields relevant and sufficient evidence for long-context retrieval without altering the underlying RAG architecture.

\subsection{Ablation Study}
To better understand the role of each design component, we conduct ablations over the three alignment strategies and the curriculum process (Table~\ref{tab:ablation}). Forward alignment proves most critical, as it directly measures chunk sufficiency for generating correct answers; its removal leads to the largest drop in performance. Backward alignment further supports complex reasoning by enforcing semantic coherence, and removing it causes retrieved passages to match superficially but lack utility. Parameter alignment, though less dominant, stabilizes training by anchoring the embedding space, reducing overfitting and collapse in noisier domains. 

The curriculum stages also show clear benefits: Stage~1 (answer-aligned positives with in-batch negatives) provides a strong initial boost, Stage~2 introduces coarse hard negatives from larger subgraphs to expose the model to more diverse distractors and enhance robustness, and Stage~3 employs fine-grained negatives from smaller subgraphs for sharper discrimination. This progressive structure teaches the retriever not only to capture relevance but also to identify evidence truly necessary for accurate answer generation.

\begin{table}[t]
\centering
\small
\setlength{\tabcolsep}{3pt}
\renewcommand{\arraystretch}{1.15}
\begin{tabular}{llllll}
\toprule
\textbf{\textsc{Method}} & \textbf{mus} & \textbf{2wiki} & \textbf{nar} & \textbf{tech} & \textbf{phil} \\
\midrule
\rowcolor{gray!15}
\multicolumn{6}{c}{\textit{\textbf{Llama3.1-8B-Instruct}}}\\
\midrule
\textbf{\textsc{Qwen3}}  & 9.63  & 28.26 & \textbf{17.70} & 19.91 & 20.72 \\
\textbf{\textsc{ARK}}    & \textbf{11.48} {\color{red}\scriptsize(+)} & \textbf{36.00} {\color{red}\scriptsize(+)} & 15.04 {\color{blue}\scriptsize(-)} & \textbf{21.47} {\color{red}\scriptsize(+)} & \textbf{21.17} {\color{red}\scriptsize(+)} \\
\midrule
\rowcolor{gray!15}
\multicolumn{6}{c}{\textbf{\textit{Qwen2.5-7B-Instruct}}}\\
\midrule
\textbf{\textsc{Qwen3}}  & 10.74 & 27.65 & 16.70 & 19.46 & 22.62 \\
\textbf{\textsc{ARK}}    & \textbf{14.61} {\color{red}\scriptsize(+)} & \textbf{29.47} {\color{red}\scriptsize(+)} & \textbf{16.77} {\color{red}\scriptsize(+)} & \textbf{20.09} {\color{red}\scriptsize(+)} & \textbf{24.64} {\color{red}\scriptsize(+)} \\
\bottomrule
\end{tabular}
\caption{Transferability of the ARK to different generators.}
\label{tab:transfer_generators}
\vskip -0.22in
\end{table}

\subsection{Transferability Across Generators}

We further investigate whether the retriever fine-tuned on \textsc{Qwen3-embedding} can transfer to other generators without additional adaptation. We evaluate end-to-end QA by directly plugging ARK into two instruction-tuned LLMs: \textsc{Llama-3.1-8B-Instruct} and \textsc{Qwen2.5-7B-Instruct}. As shown in Table~\ref{tab:transfer_generators}, ARK consistently improves performance across both generators, indicating the generalization of  out training method. A slight drop on \textsc{nar} with Llama-3.1 suggests interaction between generator decoding preferences and retrieval signals, motivating future work on generator-aware optimization or joint training aligned with generation loss.

\section{Conclusion}\label{sec:conclusion}

We propose \textsc{ARK}, a fine-tuning framework for retrievers that uses curriculum learning to integrate KG knowledge through hard-negative generation. ARK constructs a compact knowledge subgraph using LLM-generated KG and PPR-based cohesive community search; selects positive examples via three-way alignment while preserving the base encoder's similarity signal; and applies a three-stage curriculum with augmented query retrieval that incrementally incorporates harder negative chunks. The framework requires no architectural modifications, which fits standard RAG pipelines. Experiments on UltraDomain and LongBench demonstrate consistent improvements in F1 score and pairwise win-rate.

\section*{Limitations}

While our framework demonstrates strong effectiveness across diverse domains and tasks, it also has several limitations. First, our evaluation is constrained to publicly available benchmarks, which may not fully capture the diversity of real-world applications. In addition, while inference is KG-free, our training pipeline depends on an LLM-derived KG for hard-negative mining; noise in entity extraction or linking can affect the curriculum quality. We leave these aspects for future work.

\bibliography{custom}

\newpage
\appendix
\onecolumn

\section{KG-based Query Generation}

The KG serves as a core component of our framework, providing structured semantic representations that enable both entity-level reasoning and query augmentation. In this section, we detail the KG-construction, hyperparameters, and prompts used for Entity Extraction, and Query Generation.

\subsection{KG construction}

Our KG construction pipeline transforms unstructured text into a structured graph of entities and relations. Using an LLM-based extraction process followed by embedding-driven augmentation, we ensure semantic consistency and connectivity between related concepts. The algorithm \ref{alg:KG_construction} outlines this process.

\begin{algorithm*}[htbp]
\caption{KG Construction \label{alg:KG_construction}}
\DontPrintSemicolon
\KwIn{
    Context $\texttt{context}$, Chunk size $B$ and overlap size $b$, Selected generator $\texttt{LLM}$, Embedding model $\texttt{EMB}$, Similarity threshold $\tau$
}
\KwOut{Generated KG $G(V,E)$}
\BlankLine

\tcp{1. Chunking and Extraction}

$\texttt{chunks} \leftarrow \text{Chunk}(\texttt{context}, B, b)$ \Comment{Using token-based chunking} 

$\texttt{Entities, Rels} \leftarrow \{\},\{\}$ 

\ForEach{$\texttt{chunk}$ in $\texttt{chunks}$}{
    $\texttt{llm\_output} \leftarrow \texttt{LLM}(\texttt{prompt,chunk})$ 
    
    $(\texttt{extracted\_ents}, \texttt{extracted\_rels}) \leftarrow \text{ParseLLMOutput}(\texttt{llm\_output})$ 
    
    $\texttt{Entities} \cup (\texttt{ent}, \texttt{chunk\_id})$ for $\texttt{ent}$ in $\texttt{extracted\_ents}$ 
    
    $\texttt{Rels} \cup (\texttt{tuple}, \texttt{chunk\_id})$ for $\texttt{tuple}$ in $\texttt{extracted\_rels}$ 
    
}

$G \leftarrow \text{KG\_generation}(\texttt{Entities}, \texttt{Rels})$ \;   \Comment{Construct KG using LLM-generated entities and relations} 

\tcp{2. Graph Augmentation}
$\texttt{sim\_matrix} \leftarrow \text{GetCosSimilarityMatrix}(G.\text{nodes})$ \;\Comment{Using embedding model to calculate cosine similarity} 

\For{$i$ from $0$ to $\texttt{sim\_matrix}.\text{rows} - 1$}{
    \For{$j$ from $i+1$ to $\texttt{sim\_matrix}.\text{cols} - 1$}{
        \If{$\texttt{sim\_matrix}[i, j] > \tau$}{
            $\texttt{src} \leftarrow \texttt{ent}[i]$ \;
            $\texttt{tgt} \leftarrow \texttt{ent}[j]$ \;
            $\texttt{tuple} \leftarrow (\texttt{src}, \texttt{Rel\_aug}, \texttt{tgt})$ \;
            $G.\text{InsertEdge}(\texttt{tuple}, \texttt{None})$ 
        }
    }
}
\BlankLine
\Return{$G$}
\end{algorithm*}

\subsection{Hyperparmeter}

Table \ref{tab:KG_hyper} summarizes the key hyperparameters used in document processing, KG generation, and PPR retrieval. The default settings are empirically chosen to balance computational efficiency and representation quality.

\begin{table*}[htbp]
\centering
\caption{Key hyperparameters used in Query Generation. \label{tab:KG_hyper}}
\begin{tabular}{cll}
\toprule
\textbf{Category} & \textbf{Parameter} & \textbf{Default} \\
\midrule
\multirow{3}{*}{KG Construction} 
 & $\texttt{LLM}$ & Gemini-2.5 Flash \\
 & Batch size $B$ & 512 \\
 & Overlap $b$ & 12 \\
\midrule
\multirow{2}{*}{KG Augmentation} 
 & $\texttt{EMB}$ & Qwen3-Embedding 0.6B \\
 & $\tau$ & 0.8 \\
\midrule
\multirow{4}{*}{Cohesive Subgraph Search} 
 & PPR $\alpha$ & 0.85 \\
 & PPR $\epsilon$ & 1e-4 \\
 & $k_{\min}^{\text{stage2}},k_{\min}^{\text{stage3}}$ & 20, 3 \\
 & $k_{\max}^{\text{stage2}},k_{\max}^{\text{stage3}}$ & 50, 10 \\
\midrule
\multirow{1}{*}{Query Augmentation} 
 & Augmented Query size $\mathcal{Q}^{\texttt{aug}}$ & 10 \\

\bottomrule
\end{tabular}
\end{table*}

\newpage
\begingroup
\subsection{Entity Extraction Prompt}

To build a coherent KG, we use a prompt that explicitly instruct the LLM to extract entities, their attributes, and relationships in a structured format.

\begin{tcolorbox}[colback=blue!5!white, colframe=blue!75!black, breakable, title=Entity Extraction Prompt]
Given a text document that is potentially relevant to this activity, your task is to identify all entities of specific types and the relationships among them.

\textbf{Steps:}

1. Identify Entity Types: 

\quad ...

2. Extract Entities: 

\quad ...

\quad Format each entity as:

\quad \texttt{("entity"<|>entity\_name<|> \\
entity\_type<|>entity\_description>)}

3. Identify Relationships: 

\quad  ...

\quad Format each relationship as:

\quad \texttt{("relationship"\\
<|>source\_entity<|>target\_entity\\
<|>relationship\_description<|>}

\quad \texttt{relationship\_strength>)}

...

\end{tcolorbox}
\captionof{listing}{A snippet of the prompt used for entity and relationship extraction. The full prompt provides detailed instructions and examples to the LLM.}
\endgroup

\subsection{LLM Choice for KG Construction}

External LLM calls in \framework{} are used only during training to construct augmentation signals (KG/entity extraction and query rewriting) and are not required at inference time. Consequently, \framework{} introduces no additional runtime overhead once the retriever is trained.

The pipeline is model-agnostic and does not rely on proprietary systems. The required capabilities are limited to: 
(1) entity extraction and infilling, and 
(2) paraphrasing while preserving entity mentions. 
These operations do not require advanced multi-step reasoning, and reasoning-oriented models tend to increase output length and token cost without improving extraction quality. In practice, both open-source LLMs (e.g., Qwen-family) and proprietary models can be used interchangeably without architectural changes.

To quantify construction cost and graph characteristics across model choices, we report total token usage and resulting KG statistics on the first 100 samples of the \textit{Legal} subset in UltraDomain (producing 23 KGs, representing the heaviest external usage in our pipeline). We include both infilling (input) and output tokens:

\begin{table*}[h]
\centering
\small
\begin{tabular}{lccccc}
\toprule
Model & Input Tokens & Output Tokens & Avg. Nodes & Avg. Edges & Avg. Edges (Aug.) \\
\midrule
Gemini-2.5 Flash (Thinking) & 1,882,910& 15,929,992& 1144.17& 2021.83& 2935.70\\
Gemini-2.5 Flash & 1,882,910 & 2,212,372 & 788.17 & 1078.57 & 1596.65 \\
Qwen3 30B-A3B & 1,802,288 & 1,568,658 & 599.13 & 774.87 & 996.09 \\
Qwen3 235B-A22B & 1,802,288 & 1,917,941 & 675.65 & 904.26 & 1153.17 \\
\bottomrule
\end{tabular}
\caption{Token usage and resulting KG statistics under different LLM choices.}
\end{table*}

Token usage and graph statistics reveal clear differences across model choices. Enabling \textit{thinking} dramatically increases output tokens (over 7$\times$ compared to non-thinking Gemini 2.5 Flash), despite the task being structured entity-relation extraction rather than complex reasoning. Although thinking mode produces denser graphs, the structural gains are not proportional to the substantial increase in token cost, indicating that advanced reasoning is unnecessary for KG construction (if you choose a decent, large LLM model).

Among non-thinking models, output token usage remains in a similar range under identical prompts, suggesting that extraction cost is primarily governed by prompt design rather than model scale. In contrast, graph density varies more noticeably across models. Stronger models consistently generate graphs with higher average node and edge counts, indicating finer-grained entity and relation identification rather than mere verbosity. 

Overall, non-thinking LLMs, including open-source models, provide a favorable trade-off and serve as practical drop-in replacements.

\subsection{Query Generation Prompt}

Beyond KG construction, we employ a query generation process to expand the dataset and develop the curriculum learning pipeline. This prompt guides an LLM to craft semantically diverse yet answer-consistent questions by leveraging entity-level context from the KG. 

\begingroup
\begin{tcolorbox}[colback=blue!5!white, colframe=blue!75!black, title=Query Generation Prompt, breakable]

You are an expert in creating complex and confusing questions for educational purposes. Your task is to generate 10 distinct and challenging questions based on an original question-answer pair and a set of related entities with their descriptions.

The goal is to formulate questions that are semantically different from the original but lead to the exact same answer. These new questions should be confusing by design, incorporating details from the provided entity descriptions to misdirect or challenge the user's understanding.

\textbf{Input:}

You will receive the following in JSON format:

\quad - original\_question: A straightforward question.

\quad - answer: The correct and sole answer to the original question.

\quad - entities: A list of objects, where each object contains:

\quad\quad    - name: The name of the entity.
    
\quad\quad    - type: The category of the entity.
    
\quad\quad   - descriptions: A list of strings, each describing a different aspect of the entity.

\textbf{Task:}

Generate 10 new questions (we'll call them "confusing questions").

\textbf{Requirements for Confusing Questions:}

1. Same Answer: Every generated question must have the exact same answer as the original question.

2. Incorporate Entities: Each question should subtly weave in information from the entities and their descriptions.

3. Variety: The questions should be diverse in their structure and focus. You should not include exact wording / entities in the original question / answer. For example, you can: 

\quad ...

4. Clarity and Grammar: Despite being confusing, the questions must be grammatically correct and coherent.

\textbf{Output Format:}

Produce a single JSON object with one key, confusing\_questions, which contains a list of 10 string questions.

...
\end{tcolorbox}
\captionof{listing}{A snippet of the prompt used for generating confusing questions. The LLM is instructed to use provided entities to create challenging reformulations of an original question.}
\endgroup

\section{Alignment-based Finetuning}

This section describes the alignment-based finetuning procedure, which enables the model to better evaluate the quality and relevance of retrieved chunks. The alignment module computes log-likelihood scores for given question–answer pairs relative to context passages, serving as a signal for measuring faithfulness and guiding downstream retrieval calibration. Model finetuning is performed using the MS-Swift library (used primarily for fine-tuning Qwen-series model).

\begin{algorithm*}[htbp]
\caption{Three-Stage Alignment-Based Finetuning (\framework{})}
\label{alg:ark_3stage}
\DontPrintSemicolon
\SetAlgoLined
\KwIn{QA corpus $\{(q,a,\mathcal T)\}$; pretrained generator $p_\theta$; retriever sim$(\cdot,\cdot)$; weights $(\lambda_f,\lambda_b,\lambda_v)$; temperature $\tau$; augmented query sets $\mathcal Q^{\texttt{aug}}_L, \mathcal Q^{\texttt{aug}}_S$.}
\KwOut{Fine-tuned retriever parameters.}

\BlankLine
\textbf{Alignment Scoring}\;
\ForEach{$(q,a,\mathcal T)$}{
  \ForEach{$t\in\mathcal T$}{
    $S_f=\frac{1}{|a|}\sum_i \log p_\theta(a_i|q,t,a_{<i})$\;
    $S_b=\frac{1}{|q|}\sum_j \log p_\theta(q_j|a,t,q_{<j})$\;
    $S_v=\operatorname{sim}(q,t)$\;
    $S(t)=\lambda_fS_f+\lambda_bS_b+\lambda_vS_v$\;
  }
  $\mathcal T^+=\operatorname{TopM}_{t\in\mathcal T} S(t)$\;
}

\BlankLine
\textbf{Stage 1: Initial Answer-Centric Alignment}\;
\ForEach{mini-batch $\mathcal B$}{
  \ForEach{$(q,t^+)\in\mathcal B$}{
    $s(q,t)=\operatorname{sim}(q,t)/\tau$\;
    $\mathcal L_{\text{Stage1}}(q)=-\log
      \frac{e^{s(q,t^+)}}{\sum_{t'\in\text{Batch}}e^{s(q,t')}}$\;
  }
  Update retriever by $\nabla\frac{1}{|\mathcal B|}\sum\mathcal L_{\text{Stage1}}$\;
}

\BlankLine
\textbf{Stage 2: Coarse Alignment with $\mathcal Q^{\texttt{aug}}_L$}\;
\ForEach{$q'\in\mathcal Q^{\texttt{aug}}_L$}{
  Retrieve Top-$K$ chunks and form $\mathcal T^-_{\texttt{hard}_L}(q')$\;
}
\ForEach{mini-batch $\mathcal B$}{
  $\mathcal L_{\text{Stage2}}(q)=-\log
   \frac{e^{s(q,t^+)}}{\sum_{t\in\mathcal C(q)}e^{s(q,t)}}$, $\mathcal C(q)=\{t^+\}\cup\mathcal T^-_{\texttt{hard}_L}$\;
  Update retriever\;
}

\BlankLine
\textbf{Stage 3: Fine-Grained Alignment with $\mathcal Q^{\texttt{aug}}_S$}\;
\ForEach{$q''\in\mathcal Q^{\texttt{aug}}_S$}{
  Retrieve Top-$K$ chunks and form $\mathcal T^-_{\texttt{hard}_S}(q'')$\;
}
\ForEach{mini-batch $\mathcal B$}{
  $\mathcal L_{\text{Stage3}}(q)=-\log
   \frac{e^{s(q,t^+)}}{\sum_{t\in\mathcal C'(q)}e^{s(q,t)}}$, $\mathcal C'(q)=\{t^+\}\cup\mathcal T^-_{\texttt{hard}_S}$\;
  Update retriever\;
}

\Return Fine-tuned retriever\;
\end{algorithm*}

\subsection{Fintuning Pipeline}
The complete training workflow of our alignment-based retriever optimization is summarized in Algorithm \ref{alg:ark_3stage}, which details the three-stage curriculum described in Section \ref{subsec:methodology2}.

\subsection{Alignment Prompt}

The alignment prompt is designed to evaluate the faithfulness of model-generated answers relative to the provided context. In the forward alignment setting, the model is prompted to produce an answer strictly grounded in a provided context. Conversely, reverse alignment evaluates whether a given answer can be justified by its corresponding context. Together, these two directions enable a bidirectional evaluation of model reliability through log-likelihood estimation, which is later integrated into the retrieval scoring and finetuning stages.

\begingroup
\begin{tcolorbox}[colback=blue!5!white, colframe=blue!75!black, title=Forward Alignment Prompt, breakable]

You are given a context and a question. Answer the question as concisely as you can, using a single phrase or sentence if possible. If the question cannot be answered based on the information in the context, write \"unanswerable\". If the question is a yes/no question, answer \"yes\", \"no\", or \"unanswerable\". Do not provide any explanation.

\textbf{Context:} \{chunk\}

\textbf{Question:} \{question\}

\textbf{Answer:}

\end{tcolorbox}
\captionof{listing}{The forward prompt used to retrieve log-likelihood.}
\endgroup

\subsection{Hyperparameter}

Table \ref{fig:training_hyper} summarizes the key hyperparameters used during alignment-based finetuning. The process employs \textbf{full-parameter tuning} on the embedding model to ensure that the learned representations are optimally aligned with the retrieval task objectives. The configuration is chosen by default and not through grid-search since the fine-tuning stage is relatively stable.

\begin{table*}[h]
\centering
\caption{Training Hyperparameters \label{fig:training_hyper}}
\begin{tabular}{ll}
\toprule
\textbf{Parameter} & \textbf{Default} \\
\midrule
\texttt{Epochs} & 10 \\
\texttt{batch size} & 2 \\
\texttt{gradient accumulation steps} & 8 \\
\texttt{Learning rate} & 6e-6 \\
\texttt{Loss} & infonce \\
\bottomrule
\end{tabular}
\label{tab:training-hyperparams}
\end{table*}

\section{Inference}

This section describes the inference components of our system. The inference module generates final answers conditioned on those retrieved ones.

\subsection{Inference Prompt}

The inference prompt serves as the main instruction template for generating final answers from retrieved text chunks. It is intentionally concise and task-oriented, ensuring that responses are direct, factual, and free from unnecessary reasoning chains.

\begingroup
\begin{tcolorbox}[colback=blue!5!white, colframe=blue!75!black, title=Inference Prompt,breakable]

You are given a scientific article and a question. Answer the question as concisely as you can, using a single phrase or sentence if possible. If the question cannot be answered based on the information in the article, write "unanswerable". If the question is a yes/no question, answer "yes", "no", or "unanswerable". Do not provide any explanation.

\textbf{Context:} \{chunk\}

\textbf{Question:} \{question\}

\textbf{Answer:}

\end{tcolorbox}
\captionof{listing}{The general prompt used to generate answers from the retrieved context.}
\endgroup

\section{Win-rate Evaluation}
This section presents the evaluation protocol and results used to compare the performance of different retrieval and reasoning models. We adopt an LLM-based evaluator that systematically measures pairwise model performance through criteria grounded in faithfulness and conciseness. 

\subsection{Win-rate Prompt}
To ensure consistent and interpretable evaluation, we employ a structured prompt that directs an LLM to act as a neutral expert judge. The evaluator receives a ground truth reference, a question, and two candidate answers. It then performs a two-stage comparison: first applying a disqualification rule to detect unsupported answers, and subsequently assessing faithfulness (support from the ground truth), conciseness (brevity without loss of correctness) and overall winner.

\begingroup
\begin{tcolorbox}[colback=blue!5!white, colframe=blue!75!black, title=Win-rate Prompt, breakable]
You are an expert evaluator. Your task is to rigorously assess two answers to a specific question, based on a provided Ground Truth. You will use two criteria: Faithfulness and Conciseness.

...

\textbf{Evaluation Rules:}

\underline{Disqualification Rule (Primary Check):}

First, check if either answer explicitly states that the Ground Truth document does not contain enough information or evidence to answer the Question.

...

\underline{Evaluation Criteria (Secondary Check):}

\textbf{Faithfulness:} The degree to which the answer is exclusively and accurately supported by the provided Ground Truth document.

\textbf{Conciseness:} The degree to which the answer avoids mentioning excessive entities or relationships that are not essential for answering the Question.

...

\textbf{Output Format:}

Output your complete evaluation in the following JSON format.

\{

 \quad "Faithfulness": \{"Winner": "[Answer 1, Answer 2, Tie, or None]",  "Explanation": ...\},
  
  \quad  "Conciseness": \{"Winner": "[Answer 1, Answer 2, Tie, or None]", "Explanation": ...\},
  
  \quad  "Overall Winner": \{"Winner": "[Answer 1, Answer 2, Tie, or None]", "Explanation": ...\}
  
\}
\end{tcolorbox}
\captionof{listing}{A snippet of the prompt for LLM-based evaluation. The prompt defines a strict set of rules, including a disqualification rule, and requires a structured JSON output.}
\endgroup

\subsection{Additional Results}

Table \ref{tab:performance} extends win-rate comparisons to Faithfullness and Conciseness.

\begin{table*}[htbp]
  \centering
  \small
  \caption{\textbf{Additional win-rate results}. $\text{winrate} = 1/|q|\times \Sigma_q[{1_\text{Answer 1}}/({1_\text{Answer 1} + 1_\text{Answer 2}})].$ Cell shading indicates relative win rates compared to \framework.}
  \label{tab:performance}
  \resizebox{\textwidth}{!}{%
  \begin{tabular}{cccccccccccc}
    \toprule
    \multirow{2}{*}{\textsc{\textbf{Criteria}}} & \multirow{2}{*}{\textsc{\textbf{Models}}}
      & \multicolumn{5}{c}{\textbf{LongBench}}
      & \multicolumn{5}{c}{\textbf{UltraDomain}} \\
    \cmidrule(lr){3-7}  \cmidrule(lr){8-12}
    & & \textbf{nar} & \textbf{qas} & \textbf{mus} & \textbf{2wiki} & \textbf{hot}
      & \textbf{bio} & \textbf{fic} & \textbf{music} & \textbf{tech} & \textbf{phil} \\
    \midrule
    \multirow{9}{*}{\textsc{\textbf{Faithfulness}}} &
    Full
      & \scorecell{83.10}& \scorecell{58.00}& \scorecell{75.00}& \scorecell{64.44}& \scorecell{71.11}& \scorecell{100.00}& \scorecell{100.00}& \scorecell{94.12}& \scorecell{100.00}& \scorecell{100.00}\\
    \cmidrule(lr){2-12}
    & Qwen3-embedding
      & \scorecell{59.57}& \scorecell{54.17}& \scorecell{63.16}& \scorecell{61.11}& \scorecell{77.42}& \scorecell{100.00}& \scorecell{92.31}& \scorecell{94.74}& \scorecell{78.57}& \scorecell{58.82}\\
    & BGE-M3
      & \scorecell{62.26}& \scorecell{55.10}& \scorecell{63.41}& \scorecell{58.14}& \scorecell{56.76}& \scorecell{70.59}& \scorecell{91.67}& \scorecell{64.71}& \scorecell{76.92}& \scorecell{63.16}\\
    & Stella-v5
      & \scorecell{68.52}& \scorecell{84.62}& \scorecell{64.29}& \scorecell{40.00}& \scorecell{61.76}& \scorecell{72.22}& \scorecell{66.67}& \scorecell{70.59}& \scorecell{76.92}& \scorecell{89.47}\\
    & Jina-emb-v3
      & \scorecell{60.78}& \scorecell{62.50}& \scorecell{55.88}& \scorecell{53.85}& \scorecell{45.45}& \scorecell{77.78}& \scorecell{66.67}& \scorecell{66.67}& \scorecell{60.00}& \scorecell{55.56}\\
    \cmidrule(lr){2-12}
    & GraphRAG
      & \scorecell{89.13}& \scorecell{88.75}& \scorecell{77.27}& \scorecell{73.33}& \scorecell{76.74}& \scorecell{100.00}& \scorecell{100.00}& \scorecell{90.00}& \scorecell{95.00}& \scorecell{100.00}\\
    & LightRAG
      & \scorecell{76.40}& \scorecell{91.86}& \scorecell{76.71}& \scorecell{76.32}& \scorecell{89.41}& \scorecell{100.00}& \scorecell{70.00}& \scorecell{100.00}& \scorecell{100.00}& \scorecell{100.00}\\
    & HippoRAG
      & \scorecell{69.01}& \scorecell{65.00}& \scorecell{46.03}& \scorecell{40.91}& \scorecell{60.87}& \scorecell{77.78}& \scorecell{37.50}& \scorecell{50.00}& \scorecell{60.00}& \scorecell{70.00}\\
    & MemoRAG
      & \scorecell{79.17}& \scorecell{78.79}& \scorecell{68.33}& \scorecell{57.38}& \scorecell{65.38}& \scorecell{93.33}& \scorecell{93.75}& \scorecell{84.21}& \scorecell{87.50}& \scorecell{88.24}\\

    \midrule

    \multirow{9}{*}{\textsc{\textbf{Conciseness}}} &
    Full
      & \scorecell{90.24} & \scorecell{52.31} & \scorecell{81.71} & \scorecell{60.94} & \scorecell{75.00} & \scorecell{94.74} & \scorecell{100.00} & \scorecell{83.33} & \scorecell{94.74} & \scorecell{100.00} \\
    \cmidrule(lr){2-12}
    & Qwen3-embedding
      & \scorecell{55.17} & \scorecell{50.00} & \scorecell{65.31} & \scorecell{57.69} & \scorecell{62.79} & \scorecell{90.00} & \scorecell{78.57} & \scorecell{73.68} & \scorecell{76.92} & \scorecell{44.44} \\
    & BGE-M3
      & \scorecell{55.88} & \scorecell{46.97} & \scorecell{48.94} & \scorecell{56.90} & \scorecell{51.92} & \scorecell{70.59} & \scorecell{69.23} & \scorecell{38.89} & \scorecell{66.67} & \scorecell{55.00} \\
    & Stella-v5
      & \scorecell{72.46} & \scorecell{66.67} & \scorecell{61.22} & \scorecell{60.00} & \scorecell{71.11} & \scorecell{72.22} & \scorecell{62.50} & \scorecell{61.11} & \scorecell{71.43} & \scorecell{78.95} \\
    & Jina-emb-v3
      & \scorecell{62.50} & \scorecell{50.00} & \scorecell{55.10} & \scorecell{61.82} & \scorecell{50.00} & \scorecell{66.67} & \scorecell{53.85} & \scorecell{60.00} & \scorecell{58.82} & \scorecell{40.00} \\
    \cmidrule(lr){2-12}
    & GraphRAG
      & \scorecell{95.83} & \scorecell{96.63} & \scorecell{95.74} & \scorecell{95.29} & \scorecell{97.78} & \scorecell{100.00} & \scorecell{100.00} & \scorecell{90.00} & \scorecell{90.00} & \scorecell{100.00} \\
    & LightRAG
      & \scorecell{98.94} & \scorecell{96.70} & \scorecell{96.34} & \scorecell{94.19} & \scorecell{97.85} & \scorecell{100.00} & \scorecell{100.00} & \scorecell{95.00} & \scorecell{100.00} & \scorecell{100.00} \\
    & HippoRAG
      & \scorecell{79.22} & \scorecell{45.45} & \scorecell{48.57} & \scorecell{17.33} & \scorecell{48.15} & \scorecell{33.33} & \scorecell{22.22} & \scorecell{50.00} & \scorecell{60.00} & \scorecell{40.00} \\
    & MemoRAG
      & \scorecell{86.75} & \scorecell{82.19} & \scorecell{72.37} & \scorecell{67.09} & \scorecell{70.42} & \scorecell{86.67} & \scorecell{94.12} & \scorecell{84.21} & \scorecell{81.25} & \scorecell{76.47} \\
    
    \bottomrule
  \end{tabular}%
  }
  \vskip -0.1in
\end{table*}

\end{document}